\documentstyle[aps,epsf,eqsecnum]{revtex}
\begin{document}
\draft

\thispagestyle{empty}
{\baselineskip0pt
\leftline{\large\baselineskip16pt\sl\vbox to0pt{\hbox{\it Department of Physics}
               \hbox{\it Kyoto University}\vss}}
\rightline{\large\baselineskip16pt\rm\vbox to20pt{\hbox{KUNS 1614}
               \hbox{November 1999}
\vss}}%
}

\begin{center}{\large \bf 
Gravitational Radiation from a Naked Singularity. II}
\end{center}
\begin{center}{{\it --- Even-Parity Perturbation ---}}
\end{center}

\begin{center}
 {\large 
Hideo Iguchi 
\footnote{ Electronic address: iguchi@tap.scphys.kyoto-u.ac.jp}
, Tomohiro Harada
\footnote{ Electronic address: harada@tap.scphys.kyoto-u.ac.jp}\\
{\large\em Department of Physics,~Kyoto University,} \\
{\em Kyoto 606-8502,~Japan}\\
\begin{center}
 and
\end{center}
 Ken-ichi Nakao
\footnote{ Electronic address: knakao@sci.osaka-cu.ac.jp}}\\
{\large\em Department of Physics,~Osaka City University,} \\
{\large\em Osaka 558-8585,~Japan}\\
\end{center}

\begin{abstract}
A naked singularity occurs in the generic collapse
of an inhomogeneous dust ball.
We study the even-parity mode of gravitational waves from a naked
singularity of the Lema\^{\i}tre-Tolman-Bondi spacetime. 
The wave equations for gravitational waves are solved 
by numerical integration using the single null coordinate.
The result implies that the metric perturbation grows when it approaches 
the Cauchy horizon and diverges there, although
the naked singularity is not a strong source 
of even-parity gravitational radiation. 
Therefore, the Cauchy horizon in this spacetime should be 
unstable with respect to linear even-parity perturbations.
\end{abstract}

\section{Introduction}

The singularity theorems reveal that the occurrence of 
singularities is a generic property of spacetime in general 
relativity \cite{Penrose:1965wq,Hawking:1967,Hawking:1970sw}. 
However, these theorems state nothing about the detailed features of 
the singularities themselves; for example, 
we do not get information from these theorems 
about whether or not the predicted singularity is naked. 
Here, ``naked'' means that the singularity is in principle observable. 
A singularity is a boundary of spacetime. Hence, 
in order to obtain a solution of 
hyperbolic field equations for 
matter, gauge fields and spacetime itself 
in the causal future of a naked singularity, we need to impose 
a boundary condition on it. 
However, we do not yet know physically reasonable 
boundary conditions for singularities,  
and hence to avoid this difficulty, 
the cosmic censorship hypotheses (CCH) 
proposed by Penrose \cite{Penrose:1969pc,Penrose:1979} are often 
adopted in the analysis of physical phenomena involving
strong gravitational fields. 

Unfortunately no one has ever succeeded in the proof of any version of the 
CCH. There is no precise statement of CCH which can be readily proved at 
this time. Given this situation it is worth trying to obtain
counterexamples. Much effort has been made to search for 
naked singularity formation in gravitational collapse.

In the Lema\^{\i}tre-Tolman-Bondi (LTB) 
spacetime \cite{Tolman:1934,Bondi:1947},  a
naked shell-focusing singularity appears from generic initial data for
spherically symmetric configurations of the rest mass density and a
specific energy of the dust
fluid \cite{Eardley:1979tr,Christodoulou:1984,Newman:1986,Joshi:1993zg}. \
The initial functions in the most general expandable form have been
considered \cite{Jhingan:1997ia}. \ 
The matter content in this
spacetime may satisfy even the dominant energy condition.
These results are summarized as follows:
In this spacetime, a naked singularity appears from  
generic initial data for spherically symmetric 
configurations of the rest mass density and a specific energy 
of the dust fluid. 
Shapiro and Teukolsky numerically studied evolution of 
collisionless gas spheroids with fully general relativistic 
simulations \cite{Shapiro}. 
They found some evidence that prolate spheroids 
with sufficiently elongated initial configurations, and 
even with some angular momentum, may form naked singularities.
Ori and Piran numerically examined the structure of self-similar spherical
collapse solutions for a perfect fluid with a barotropic equation of
state \cite{Ori:1987hg,Ori:1990ps}. They showed that there is a globally naked
singularity in a significant part of the space of self-similar
solutions. Joshi and Dwivedi analytically investigated the self-similar 
spherically symmetric collapse of a perfect fluid with a
similar equation of state \cite{Joshi:1992}. Harada numerically investigated 
spherical collapse of a perfect fluid
without the assumption of self-similarity \cite{Harada:1998wb}. 
A spherical cloud of counterrotating particles was investigated by
the present authors \cite{Harada:1998cq}.
The spherical gravitational collapse of an imperfect fluid which has
only a tangential pressure has also been considered \cite{Magli:1997,Magli:1998,Singh:1997iy,Barve:1999ph,Harada:1999gg}.
Further, the naked singularity produced by the gravitational 
collapse of radiation shells \cite{JoshiIII} 
and of more general matter \cite{Dwivedi:1994qs} were investigated. 
As for the non-spherically symmetric collapse case, 
Joshi and Krolak revealed that a naked singularity 
appears also in the Szekeres spacetime 
with irrotational dust matter \cite{Joshi:1996qc}.
The global visibility of this singularity was recently analyzed \cite{Deshingkar:1998ge}.

In this paper we investigate whether a naked singularity, if such
exists, is a strong source of gravitational radiation, 
and we attempt to understand the dynamics and 
observational meaning of the naked singularity formation.
As noted above, several researchers have shown that the final fate of
gravitational collapse is not always a singularity covered by an event
horizon. 
In this case with a small disturbance of spacetime, very short
wavelength gravitational waves, which are  
created in the high density region around a singularity,
may propagate to the observer outside the dust cloud because of the
absence of an event horizon. If this is true,  
extremely high energy phenomena which cannot be realized in
any high energy experiment on Earth can be observed. 
Moreover, information regarding the physics of so-called `quantum gravity'
may be obtained. Also, these waves may be
so intense that they destroy the Cauchy horizon.
In this paper we consider  the generation of gravitational waves 
during the collapse of a spherical dust ball with a small disturbance 
of the density profile, i.e. perturbations of LTB spacetime.

Nakamura, Shibata and Nakao \cite{Nakamura:1993xr} have suggested that a 
naked singularity may emit considerable gravitational wave radiation.
This was proposed using an estimate of gravitational radiation from a
spindle-like naked singularity.
They modeled the spindle-like 
naked singularity formation in gravitational collapse
using a sequence of general relativistic, momentarily static initial data
for a prolate spheroid.
It should be noted that the system they
considered is different from that considered in this
article and that their result is controversial. There
are numerical analyses that may support or may not support the results of 
Nakamura, Shibata and
Nakao for prolate collapse \cite{Shapiro} and for cylindrical 
collapse \cite{Echeverria:1993wf,Chiba:1996}.

Due to the non-linear nature of the problem, 
it is difficult to analytically solve the Einstein equation. Therefore, 
numerical methods will provide the final tool. However, 
its singular behavior makes 
accurate numerical analysis very difficult at some stage. 
In this article, we investigate even-parity linear 
gravitational waves from the collapse of an inhomogeneous
spherically symmetric dust cloud. 
 Even for the linearized Einstein equation we must perform 
numerical integration. 
However, in contrast to the numerical simulation of 
the full Einstein equation, 
high precision is guaranteed for the numerical integration of
the linearized Einstein
equation, even in regions with extremely large spacetime curvature. 
Furthermore, the linear stability of known examples 
of naked singularity formation is necessary 
as a first step to understand the general dynamics near naked 
singularity formation.

Recently, Iguchi, Nakao and Harada 
\cite{Iguchi:1998qn} (INH) studied odd-parity metric
perturbations around a naked singularity in the LTB spacetime. In INH,
it was found that the propagation of odd-parity gravitational waves is
not affected by the collapse of a dust cloud before the formation of the event
horizon, even if there appears a
central naked singularity. The same authors extended their study
to consider the
generation of gravitational waves from the dust collapse including 
matter perturbation \cite{Iguchi:1999ud}.\ They showed that
gauge-invariant variables diverge only at the center, and they do not
propagate to  
the outside. For an odd-parity perturbation the evolution of the matter
perturbation decouples from the evolution of the metric
perturbation, while the even-parity matter perturbation couples to the metric 
part. Therefore an even mode seems to be more essential. To investigate 
the generation of gravitational waves in LTB spacetime we should analyze 
even-parity perturbations.  Here we investigate the
behavior of the even-parity quadrupole metric and matter perturbations
in the marginally bound LTB background. We numerically
calculate the time evolutions of the gauge invariant metric
variables. We show that some of metric perturbation variables and the Weyl
scalar diverge at the Cauchy horizon but that derived the energy flux does not.

This paper is organized as follows: In Sec. \ref{sec:be} the basic
equations are derived; in Sec. \ref{sec:results} the numerical
results are presented; in Sec. \ref{sec:discussion} we discuss the 
numerical results; and in Sec. \ref{sec:summary} we summarize our
results. We adopt geometrized units in which  $c=G=1$. 
The signature of the metric tensor and 
sign convention for the Riemann tensor follow 
Ref.\ \cite{MTW}.

\section{Basic equations}
\label{sec:be}
We consider the evolution of even-parity perturbations of the LTB
spacetime to linear order. The background LTB spacetime describes
the dynamics of an inhomogeneous spherically symmetric dust ball.
Using the synchronous comoving coordinate system, the 
line element of the LTB spacetime can be expressed in the form  
\begin{equation}
  \label{bgmetric}
  d{\bar s}^{2}= {\bar g}_{\mu\nu}dx^{\mu}dx^{\nu}\equiv
  -dt^2+A^{2}(t,r)dr^{2}
  +R^2(t,r)(d\theta^{2}+\sin^{2}\theta d\phi^{2}).
\end{equation}
The energy-momentum tensor for the dust fluid is 
\begin{equation}
  \label{bgmatter}
  {\bar T}^{\mu\nu} = {\bar \rho}(t,r){\bar u}^{\mu}{\bar u}^{\nu},
\end{equation}
where ${\bar\rho}(t,r)$ is the rest mass density and 
${\bar u}^{\mu}$ is the 4-velocity of the dust fluid. 
In the synchronous coordinate system, the unit vector field  normal to 
the spacelike hypersurfaces is geodesic, and  
there is a freedom concerning which timelike geodesic field 
is adopted as the hypersurface unit normal. 
Using this freedom, we can always set ${\bar u}^{\mu}=\delta^{\mu}_{0},$ 
since the 4-velocity of the spherically symmetric 
dust fluid is tangent to an irrotational 
timelike geodesic field. 

Then the Einstein equations and the equation of motion for the 
dust fluid reduce to the following simple equations:
\begin{eqnarray}
  A &=&  \frac{R'}{\sqrt{1+f(r)}}, \label{eq:A} \\
  {\bar \rho}(t,r) &=& \frac{1}{8\pi}
  \frac{1}{R^2 R'}{dF(r)\over dr},\label{eq:einstein} \\
  \dot{R}^2-\frac{F(r)}{R} &=& f(r).\label{eq:energyeq} 
\end{eqnarray}
Here $f(r)$ and $F(r)$ are arbitrary functions of the radial 
coordinate, $r$, and the overdot and prime denote partial derivatives
with respect to $t$ and $r$, respectively. From Eq.\ (\ref{eq:einstein}), $F(r)$ is related 
to the Misner-Sharp mass function \cite{Misner}, $m(r)$, 
of the dust cloud in the manner 
\begin{equation}
  \label{mass}
  m(r) = 4\pi \int_0^{R(t,r)}{\bar \rho}(t,r)R^2dR = 4\pi
  \int_0^r{\bar \rho}(t,r)R^2R' dr =\frac{F(r)}{2}. 
\end{equation}
Hence Eq.\ (\ref{eq:energyeq}) might be 
regarded as the energy equation per unit mass. 
This means that  
the other arbitrary function, $f(r)$, is recognized as 
the specific energy of the dust fluid. 
The motion of the dust cloud is completely specified 
by the function $F(r)$ 
(or equivalently, the initial distribution of 
the rest mass density, ${\bar \rho}$) and the specific energy, $f(r)$. 
When we restrict our calculation to the case that the symmetric center, 
$r=0$, is initially regular, the central shell focusing singularity 
is naked if and only if 
$\partial_{r}^{2}{\bar\rho}|_{r=0}<0$ is initially 
satisfied for the marginally bound collapse, $f(r)=0$
\cite{Singh:1996tb,Jhingan:1996jb}. \  
For collapse that is not marginally bound, there exists a similar
condition as an inequality for a value depending on the functional forms 
of $F(r)$ and $f(r)$ \cite{Newman:1986,Singh:1996tb,Jhingan:1996jb}. \

Next we give a brief introduction to the gauge-invariant 
formalism of Gerlach and
Sengupta \cite{Gerlach:1979rw,Gerlach:1980tx} for even-parity perturbations 
around the most general spherically symmetric spacetime.
We consider the general spherically symmetric spacetime with the metric
\begin{equation}
  g_{\mu \nu}dx^{\mu}dx^{\nu} \equiv g_{ab}(x^d)dx^a dx^b
  +R^2(x^d)\gamma_{AB} (x^D)dx^A dx^B,
\end{equation}
and stress-energy tensor
\begin{equation}
  t_{\mu \nu}dx^{\mu}dx^{\nu} \equiv t_{ab}(x^d)dx^a dx^b
  +\frac{1}{2}t_A^{~A}R^2(x^d)\gamma_{AB} (x^D)dx^A dx^B
\end{equation}
where $\gamma_{AB} dx^A dx^B = d\theta ^2 + \sin ^2 \theta d \phi ^2$. 
Here, lower-case Latin indices refer to radial and time coordinate, while
capital Latin indices refer to $\theta$ and $\phi$.

The even-parity perturbations are 
\begin{eqnarray} 
  \label{GS-pmetric} 
  h_{\mu\nu}  &=& \left(\begin{array}{cc} 
                        h_{ab}(x^d)Y & h_a (x^d) Y_{:B} \\ 
                       \mbox{sym}   & K(x^d) R^2 \gamma_{AB}Y + G(x^d) R^2 Z_{AB}
                      \end{array}  \right) 
\end{eqnarray} 
for metric and
\begin{eqnarray} 
  \label{GS-pmatter} 
  \delta T_{\mu\nu} &=& \left(\begin{array}{cc} 
                        \Delta t_{ab}(x^d)Y & \Delta t_a (x^d) Y_{:B} \\ 
                        \mbox{sym} & \Delta t^3 (x^d) R^2 \gamma_{AB}Y 
+ \Delta t^2(x^d) Z_{AB} 
                      \end{array}  \right)  
\end{eqnarray} 
for matter, where $Y \equiv Y_l^m(x^D)$ are the scalar spherical
harmonics and $Z_{AB} = Y_{:AB} + \frac{l(l+1)}{2} Y \gamma_{AB}$.
Here covariant derivatives are distinguished as follows:
\begin{equation}
  \gamma_{AB:C} \equiv 0, ~~~~~ g_{ab|c}\equiv 0.
\end{equation}
For convenience of expression, we introduce 
\begin{equation}
  v_a \equiv R_{,a}/R 
\end{equation}
and
\begin{equation}
  p_a \equiv h_a - \frac{1}{2}R^2 G_{,a}.
\end{equation}
A set of even-parity gauge-invariant metric perturbations is defined as
\begin{eqnarray}
  \label{gi-kab}
  k_{ab} &\equiv& h_{ab}-(p_{a|b}+p_{b|a}) \\
  \label{gi-k}
  k &\equiv& K + \frac{l(l+1)}{2}G - 2v^a p_a.
\end{eqnarray}
A set of even-parity gauge-invariant matter perturbations is defined as
\begin{eqnarray}
  T_{ab} &\equiv& \Delta t_{ab}- t_{ab|c}p^c -t_a^{~c}p_{c|b} 
  -t_b^{~c}p_{c|a}, \\
  T_a &\equiv& \Delta t_a -t_a^{~c}p_c -R^2(t_A^{~A}/4)G_{,a},\\
  T^3 &\equiv& \Delta t^3 -(p^c/R^2)(R^2 t_A^{~A}/2)_{,c} 
  +l(l+1)(t_A^{~A}/4)G,\\
  T^2 &\equiv&  \Delta t^2 -(R^2 t_A^{~A}/2)G.
\end{eqnarray}
The perturbed Einstein equations are expressed only in gauge-invariant
perturbations as 
Eqs. (3.13) of Ref.\ \cite{Gerlach:1980tx}. We give these equations in
Appendix \ref{sec:linear}.

In this paper we restrict our numerical investigation to the 
quadrupole mode in the marginally bound background. We derive the
perturbed equations in that case. 
Note that, from Eq.\ (\ref{eq:A}), the background metric variable
 $A$ is equal to $R'$.
Also, we can easily integrate Eq.\ (\ref{eq:energyeq}) and obtain 
\begin{equation}
  \label{f=0}
  R(t,r) = \left(\frac{9F}{4}\right)^{1/3}[t_0(r)-t]^{2/3},
\end{equation}
where $t_0(r)$ is an arbitrary function of 
$r$. The formation time of the naked singularity is $t_0=t_0(0)$. 
Using the freedom
for the scaling of $r$, we choose $R(0,r)=r$. This scaling of $r$ 
corresponds to the following choice of $t_{0}(r)$: 
\begin{equation}
  \label{t0}
  t_0(r) = \frac{2}{3\sqrt{F}}r^{3/2}.
\end{equation}

The energy density $\bar{\rho}$ is perturbed by adding the scalar term $\delta
\rho Y$, while the 4-velocity $\bar{u}_\mu$ is perturbed by adding the term
\begin{equation}
  \delta u_\mu =(V_0(x^d)Y,V_1(x^d)Y,V_2(x^d)Y_{,A}).
\end{equation}
The normalization for the 4-velocity yields the relation $\bar{u}^\mu  \delta
u_\mu =0$. This relation implies that  $V_0$ vanishes exactly. 
Then there are only three matter perturbation variables, 
\begin{eqnarray}
  T_{00} &=& \delta \rho(t,r), \\
  T_{01} &=& \bar{\rho} V_1(t,r), \\
  T_0 &=& \bar{\rho} V_2(t,r).
\end{eqnarray}
The others exactly vanish:
\begin{equation}
  T_{11}=T_1=T^3=T^2=0.
\end{equation}
Now we can write
down the perturbed Einstein field equations for the background LTB
spacetime. The resulting linearized Einstein equations are given in
Appendix \ref{sec:linear}. 

We have obtained seven differential equations, (\ref{00})--(\ref{z22}),
for seven variables (four metric and three matter). The right-hand sides
of four of these equations vanish exactly. Then we can obtain the
behavior of the metric variables through the integration of them. We transform
these equations into more favorable forms.
  From Eq.\ (\ref{le4}),  
\begin{equation}
  k_{00} = \frac{1}{R'^2}k_{11}.
\end{equation}
Using this relation and the remaining equations whose r.h.s. 
vanish, we obtain evolution
equations for gauge-invariant metric variables as
\begin{eqnarray}
\label{evQ}
  -\ddot{q}+\frac{1}{R'^2}q''&=& {\frac{4}{{{R}^2}}} q 
                             + \left(\frac{2}{RR'} + \frac{R''}{R'^3}\right)q'
                             + 3\frac{\dot{R'}}{R'}\dot{q} 
                             + 4\left({\frac{\dot{R}}{R}}
                             - \frac{\dot{R'}}{R'}\right)\dot{k} \nonumber\\
                         & & + \frac{2}{R'^3}\left(- \dot{R''}
                             - {\frac{2{{R'}^2}\dot{R}}{{{R}^2}}} 
                             - {\frac{R''\dot{R}}{R}} 
                             + {\frac{2R'\dot{R'}}{R}} 
                             + {\frac{2R''\dot{R'}}{R'}} 
                            \right)k_{01}\nonumber \\
                         & & + \frac{2}{R'^3}\left(- \dot{R'} 
                             + {\frac{R'\dot{R}}{R}}\right){k_{01}}' ,
\end{eqnarray}
\begin{eqnarray}
\label{evK}
 \ddot{k}&=&    - {\frac{2}{R^2}}q
                - \frac{q'}{RR'}+ \frac{\dot{R}}{R}\dot{q}
                - 4\frac{\dot{R}}{R}\dot{k}
                + \frac{2}{RR'}\left(- \frac{\dot{R'}}{R'}
                + \frac{\dot{R}}{R}\right) {k_{01}},
\end{eqnarray}
\begin{equation}
\label{evk01}
  \dot{k_{01}} = -\frac{\dot{R'}}{R'}k_{01}-q',
\end{equation}
where $q \equiv k-k_{00}$. If we solve these three equations for some 
initial data and for the appropriate boundary conditions, we can follow 
the full evolution of the metric perturbations. When we substitute
these metric perturbations into Eqs.\ (\ref{00}), (\ref{01}) and
(\ref{02}), the matter perturbation variables $\delta
\rho, V_1$ and $V_2$, respectively, are obtained.

We can also investigate the evolution of the matter perturbations 
from the linearized conservation equations $\delta (T^{\mu \nu}_{~~;\nu})=0$.
They reduce to
\begin{eqnarray}
 \label{rhodot}
  \left(\frac{\delta \rho}{\bar{\rho}}\right)^{.} 
    &=&\frac{1}{\bar{\rho}R^2R'}\left(\frac{R^2\bar{\rho}}{R'}
       \left(k_{01}+V_1\right)\right)' -\frac{6}{R^2}V_2
       -\dot{k}-\frac{3}{2}\left(\dot{k}-\dot{q}\right),\\
 \label{V1dot}
  \dot{V_1} &=& -\frac{1}{2}\left(k'-q'\right), \\
 \label{V2dot}
  \dot{V_2} &=& -\frac{1}{2}\left(k-q\right) .
\end{eqnarray}
Integration of these equations gives us the time evolution of the matter 
perturbations. 
We can check the consistency of the numerical calculation by 
comparison of these variables and those obtained
from Eqs.\ (\ref{00}), (\ref{01}) and (\ref{02}).

To constrain the boundary conditions in our numerical calculation, we
should consider the regularity conditions at the center. These
conditions are obtained from requiring that all tensor quantities
be expandable in non-negative integer powers of locally Cartesian
coordinates near the center \cite{Bardeen}.\ The detailed derivation
of these conditions is too complicated to be presented here.
 We simply quote the results. 
The regularity conditions for the metric perturbations 
are 
\begin{equation}
  k \sim k_0(t)r^2,~~ q \sim q_0(t)r^4,~~ k_{01} \sim k_0(t)r^3.
\end{equation}
For the matter perturbations, the regularity conditions at the center
are 
\begin{equation}
  \delta \rho \sim \delta \rho_0(t) r^2,~~ V_1 \sim V_{10}(t)r,
  ~~ V_2 \sim V_{20}(t)r^2.
\end{equation}
Therefore all the variables we need to calculate vanish at the
center.

\section{Numerical method and results}
\label{sec:results}
We numerically solved the wave equations
(\ref{evQ})--(\ref{evk01}). Following the method of previous papers, 
\cite{Iguchi:1998qn,Iguchi:1999ud} we
transformed the wave equation (\ref{evQ}) into the out-going single-null
coordinate system. 
In this section, we present this coordinate transformation and explain
our background and initial data of the perturbations. In the later half
of this section, we give our numerical results.

\subsection{Numerical method}
\label{sec:method}

In the previous section it was shown
that the perturbation variables $q, k$ and $k_{01}$ vanish at the
center. A careful treatment of the differential equations may be
required near the center for proper propagation through the center.  
Hence we define the new variables 
\begin{equation}
  \tilde{q}=qR'^7/R^4,~~~ \tilde{k}=kR'^4/R^2,~~~
  \tilde{k}_{01}=k_{01}R'^5/R^3 .
\end{equation}
These new variables are not identically zero at the regular 
center and do not diverge when they approach the central singularity 
because of the suppression factor $R'$. We rewrite Eqs.\
(\ref{evQ})--(\ref{evk01}) in terms of these new variables.

Next we perform a coordinate transformation for Eq.\ (\ref{evQ})
from the synchronous comoving coordinate system $(t,r)$ to the
single-null coordinate system $(u,\tilde{r})$, where $u$ is the outgoing null
coordinate and $\tilde{r}=r$. 
We perform the numerical integration of this equation along two
characteristic directions. Therefore we use a double null grid in the
numerical calculation.  Whereas we integrate Eqs.\ (\ref{evK}) and
(\ref{evk01}) along the direction $r=\mbox{const}$.
(Detailed explanations of the single-null coordinate used in our
calculation is given in INH.) \ As a result, we obtain
the first order differential equations 
\begin{equation}
 \label{dX}
  \frac{1}{\alpha}\frac{d}{du}X=a_1 X+a_2 W+a_3 Z +a_4 \tilde{k} +a_5 \tilde{q},
\end{equation}
\begin{equation}
 \label{dotW}
  \dot{W}=b_1 X+b_2 W+b_3 Z+b_4  \tilde{k} +b_5 \tilde{q} ,
\end{equation}
\begin{equation}
 \label{dotZ}
  \dot{Z}=c_1 X+c_2 W+c_3 Z+c_4  \tilde{k} +c_5 \tilde{q} ,
\end{equation}
\begin{equation}
 \label{dotk}
  \dot{\tilde{k}}=d_1 X+d_2 W+d_3 Z+d_4  \tilde{k} +d_5 \tilde{q} ,
\end{equation}
\begin{equation}
 \label{delq}
  \partial_{\tilde{r}}\tilde{q}=e_1 X+e_2 W+e_3 Z +e_4 \tilde{k} 
  +e_5 \tilde{q},
\end{equation}
where we have introduced $X$ and $W$, which are defined by 
Eqs.\ (\ref{delq}) and (\ref{dotk}), respectively, and
\begin{equation}
  Z \equiv \tilde{k}_{01}-\frac{R}{R'}\tilde{q}.
\end{equation}
$\alpha$ is given by
\begin{equation}
  \alpha \equiv \frac{1}{\dot{u}}.
\end{equation}
The derivatives in Eqs.\ (\ref{dX}) and (\ref{delq}) are given by
\begin{eqnarray}
  \frac{d}{du} &=& \partial_u +
  \frac{d\tilde{r}}{du}\partial_{\tilde{r}}
   = \partial_u-\frac{\alpha}{2R'}\partial_{\tilde{r}} 
  =\frac{\alpha}{2}\partial_t-\frac{\alpha}{2R'}\partial_{r},\\
  \label{r'a}
  \partial_{\tilde{r}} &=& -\frac{u'}{\dot{u}}\partial_{t}+\partial_{r}=
    R'\partial_{t}+\partial_{r}.
\end{eqnarray}
The coefficients $a_1, a_2, \cdots$ are shown 
in Appendix \ref{sec:evolve}. 
Equations (\ref{dX}) and (\ref{delq}) are integrated along
the double-null grid. 
We integrate Eq.\ (\ref{dX}) using the scheme of an explicit first order 
difference equation, and we use the trapezoidal rule to integrate Eq.\
(\ref{delq}). 
Equations (\ref{dotW})--(\ref{dotk}) are integrated along the 
timelike directions $r=\mbox{const}$ using a 
first order difference method. We interpolate variables to estimate the
right-hand sides of Eqs.\ (\ref{dotW})--(\ref{dotk}) at the same radial
coordinate $r$ on the previous out-going null slice.

We adopt the initial rest mass density profile 
\begin{equation}
  \label{density}
  \rho(r)=\rho_0 \frac{1+ \exp\left(-\frac{1}{2}\frac{r_1}{r_2}\right)}
  {1+ \exp\left(\frac{r^n -r_1^n}{2 r_1^{n-1}r_2}\right)},
\end{equation}
where $\rho_0$, $r_1$ and $r_2$ are positive constants, and $n$ is a 
positive even
integer. As a result the dust fluid spreads all over the space. However,
if $r \gg r_1,r_2$, then $\rho(r)$ decreases exponentially, so that the dust
cloud is divided between the dense core region  and the envelope,
which can be
considered as the vacuum region. We define a core radius 
\begin{equation}
  \label{cradious}
  r_{\mbox{\scriptsize core}}=r_1+\frac{r_2}{2}.
\end{equation}
If we set $n=2$, there appears a central naked
singularity. This singularity becomes locally or globally naked, depending 
on the parameters $\rho_0$, $r_1$ and $r_2$. However, if the integer
$n$ is greater than $2$, the final state of the dust cloud is a black
hole for all parameter values. Then we consider three different
density profiles connected with three types of the final state of the
dust cloud, globally and locally naked singularities and a black
hole. The outgoing null coordinate $u$ is chosen so that it agrees
with the proper time at the symmetric center. 
Corresponding parameters are given in Table \ref{tab:parameter}.
Using this density profile, we numerically calculated the total 
gravitational mass of 
the dust cloud $M$. In our calculation we adopted the total
mass $M$ as the unit of the variables.

We give the numerical results from 
the initial conditions for the perturbations  
\begin{eqnarray}
  X&=&\frac{\partial_{\tilde{r}}\tilde{q}-(e_3 Z +e_5 \tilde{q})}{e_1}, \\
  W&=&-d_5 \tilde{q}, \\
  Z&=&4\frac{R}{\dot{R}R'}\tilde{q}, \\
  \tilde{k}&=&-\frac{(3R'b_1+b_5 -b_2 d_5)\tilde{q}}{b_4}, \\
  \tilde{q}&=&\left(1+\left(\frac{r}{r_3}\right)^2\right)^{-\frac{5}{2}}\frac{R^2}{\dot{R}^2}b_4, 
\end{eqnarray}
on the initial null surface. Here 
$\dot{\tilde{k}}$ vanishes on this surface and
$\dot{W}$ and $\dot{Z}$ are diminished near the center.
We chose $r_3=0.3 r_{\mbox{\scriptsize core}}$.
The main results of our numerical investigation do
not depend on the detailed choice of the initial conditions.

 \begin{table}[tb]
   \begin{center}     
 \caption{Parameters of initial density profiles and damped oscillation frequencies, where $M=1$.}
 \label{tab:parameter}
 \begin{tabular}{ccccccc}
  & final state & $\rho_0$ & $r_1$ & $r_2$ & $n$ &  damped oscillation frequency  \\ \hline
 (a) & globally naked & $1 \times 10^{-2}$ & 0.25 & 0.5 & 2 & ---    \\ 
 (b) & locally naked & $1 \times 10^{-1}$ & 0.25 & 0.5 & 2 & 0.36+0.096$i$    \\ 
 (c) & black hole & $2 \times 10^{-2}$ & 2 & 0.4 & 4 & 0.36+0.093$i$   \\ 
 \end{tabular}
   \end{center}
 \end{table}

\subsection{Results}
\label{sec:subresults}
\begin{figure}[b]
  \begin{center}
    \leavevmode
    \epsfysize=300pt\epsfbox{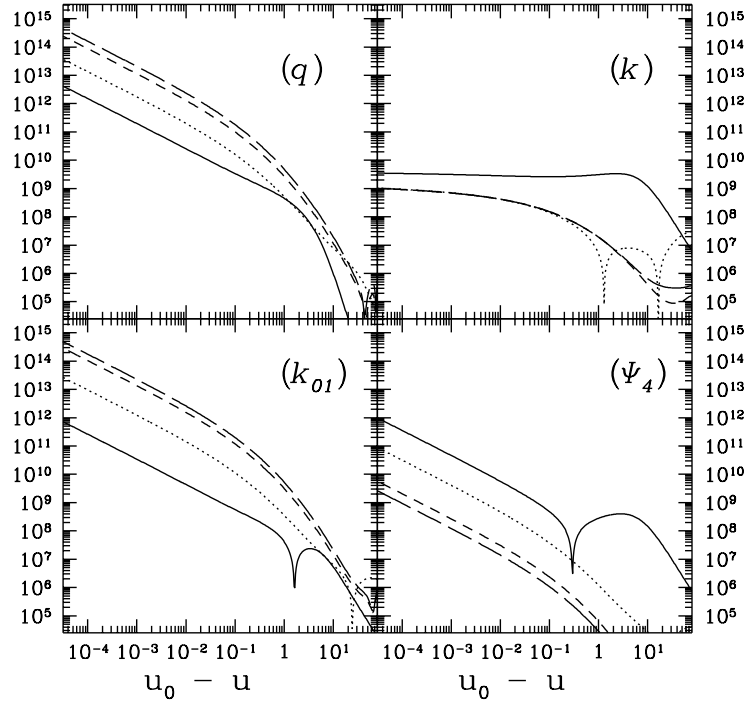}    
  \caption{Plots of perturbed variables $q, ~k, ~k_{01}$ and the Weyl scalar 
    $\Psi_4$ 
   at constant circumferential radius $R$.
    The results for $R=1$,  $R=10$, $R=100$, and $R=200$ are
    plotted. The solid lines represent the results for $R=1$, the dotted
    lines for $R=10$, the dashed lines for $R=100$, and the long dashed lines
    for $R=200$. $u=u_0$ corresponds to the Cauchy horizon. } 
 \label{fig:out}
  \end{center}
 \end{figure}

First we observe the behavior of the metric variables $q, k, k_{01}$ and 
the Weyl scalar, which corresponds to out-going waves,
\begin{eqnarray}
   \Psi_4 &\equiv& C_{\mu\nu\rho\sigma}n^{\mu}\bar{m}^{\nu}n^{\rho}\bar{m}^{\sigma}\\
     &=& -\frac{3}{32}\sqrt{\frac{5}{\pi}}\sin^2\theta\frac{k_{01}-\left(k-q\right)R'}{R^2 R'},
\end{eqnarray}
where
\begin{eqnarray}
  n^{\mu} &=& \left(\frac{1}{2},-\frac{1}{2R'},0,0\right)\\
  \bar{m}^{\nu} &=& \left(0,0,\frac{1}{\sqrt{2}R},-\frac{i}{\sqrt{2}R\sin\theta}\right),
\end{eqnarray}
outside the dust cloud. The results are plotted in
Fig.\ \ref{fig:out}. We can see that the metric variables $q, k_{01}$ and 
the Weyl scalar $\Psi_4$ diverge when they approach the Cauchy horizon. The
asymptotic power indices of these quantities are about $\sim$ 0.88. On the
other hand the metric quantity $k$ does not diverge when it approaches the
Cauchy horizon. The energy flux is computed  by constructing the
Landau-Lifshitz pseudotensor. We can calculate the radiated power of
gravitational waves from
this. The result is given in Appendix \ref{sec:power}. For the
quadrupole mode, the total radiated power becomes 
\begin{equation}
  P = \frac{3}{8\pi} k^2.
\end{equation}
The radiated power of the gravitational waves is proportional to the
square of $k$. 
Therefore the system of
spherical dust collapse with linear perturbations 
cannot be expected as a strong source of gravitational waves.

\begin{figure}[b]
  \begin{center}
    \leavevmode
    \epsfysize=400pt\epsfbox{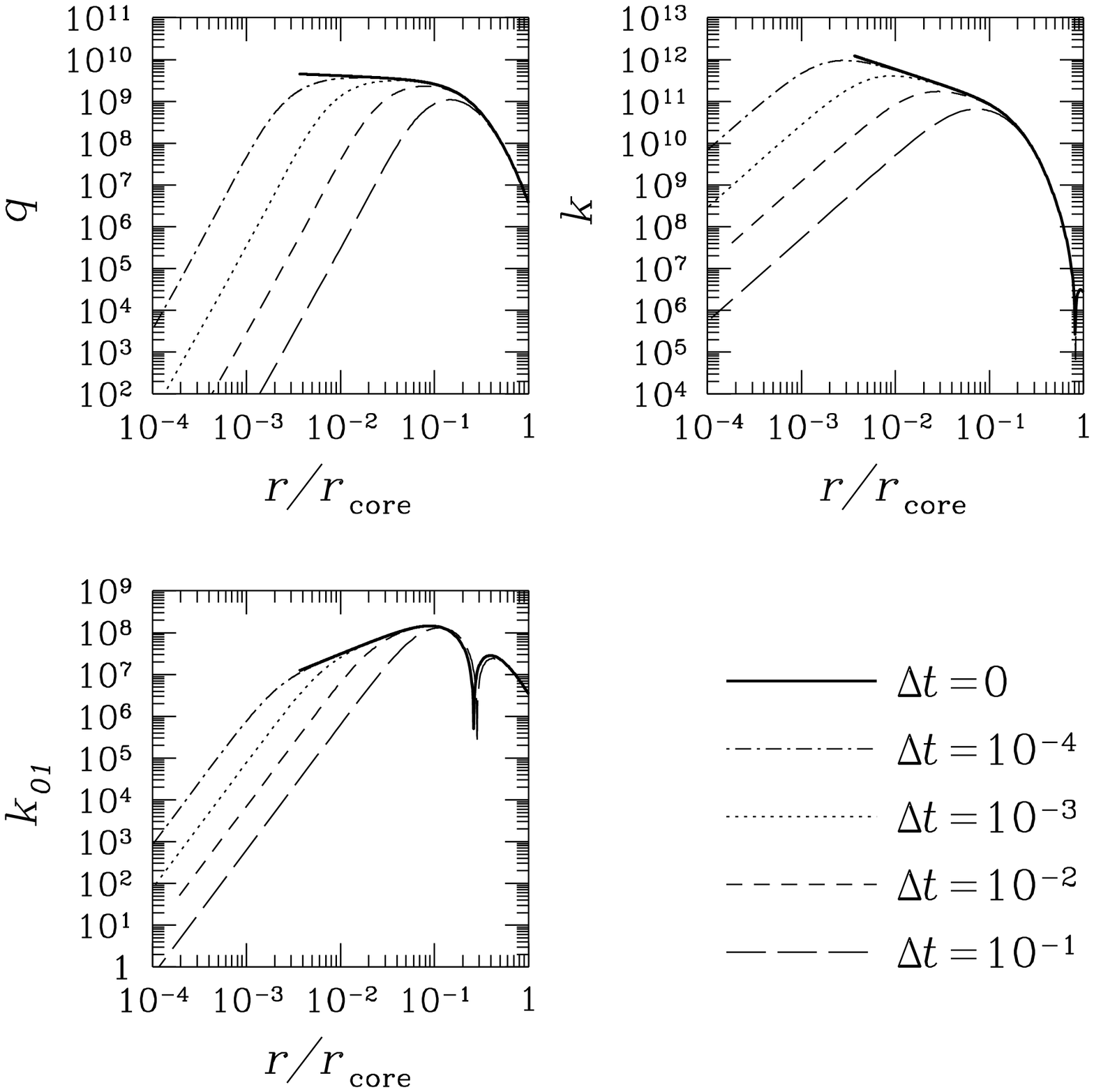}    
  \caption{Plots of perturbed variables  $q, ~k$ and $k_{01}$ near the
    center. The values for $\Delta t=t_0 -t=10^{-1}, 10^{-2}, 10^{-3},
    10^{-4}, 0$ are plotted. The solid lines represent the results for
    $\Delta t =0$, the long dashed lines for $\Delta t =10^{-1}$, the dashed 
    lines for $\Delta t =10^{-2}$, the dotted lines for $\Delta t =10^{-3}$,
    and the dotted dashed lines for $\Delta t =10^{-4}$.}
 \label{fig:center}
  \end{center}
 \end{figure}

\begin{figure}[t]
  \begin{center}
    \leavevmode
    \epsfysize=400pt\epsfbox{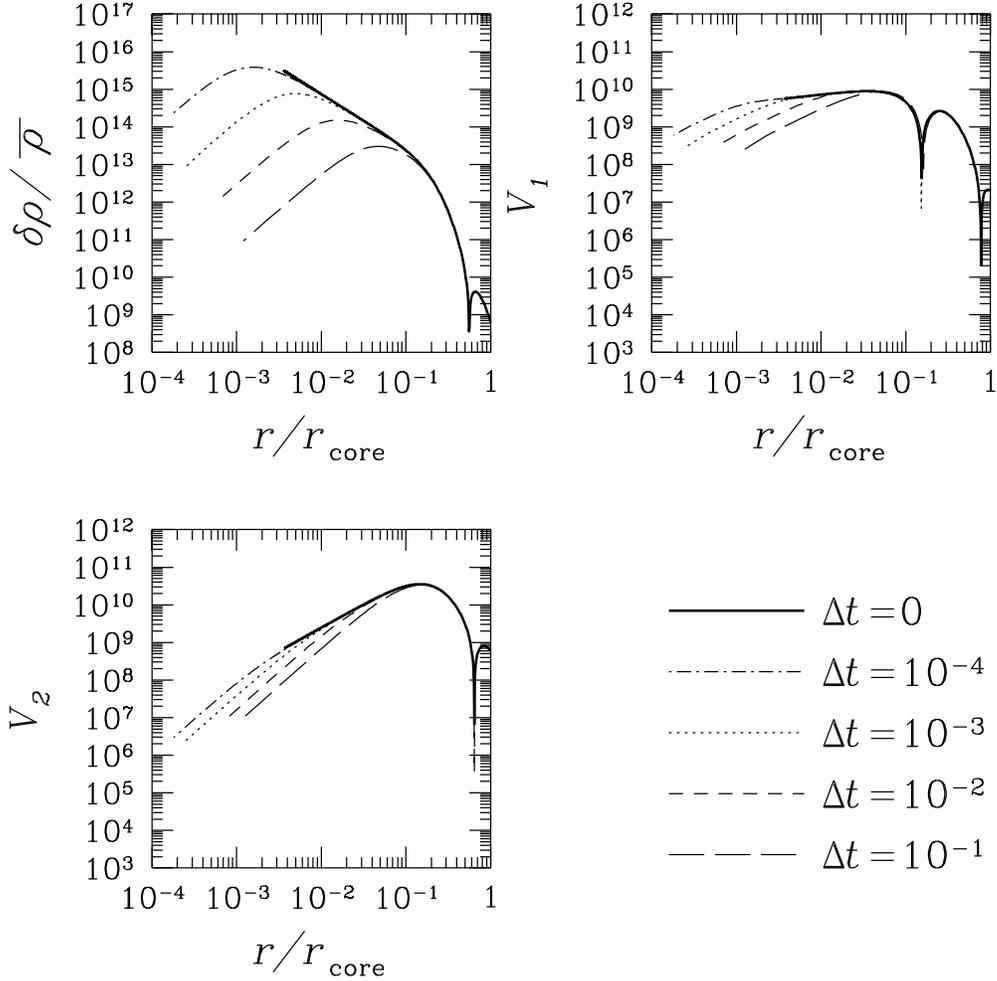}    
  \caption{Plots of perturbed variables  $\delta \rho,
    V_1,$ and $V_2 $ near the center. The values for $\Delta t =10^{-1},
    10^{-2}, 10^{-3}, 10^{-4},$ and  $0$ are
    plotted. The solid lines represent the results for
    $\Delta t =0$, the long dashed lines for $\Delta t =10^{-1}$, the dashed 
    lines for $\Delta t =10^{-2}$, the dotted lines for $ \Delta t =10^{-3}$,
    and the dotted dashed lines for $\Delta t =10^{-4}$.}
 \label{fig:centerm}
  \end{center}
 \end{figure}

Second we observe the perturbations near the center. The results 
are plotted in Figs.\ \ref{fig:center} and \ref{fig:centerm}. In these
figures we plot the perturbations at $t-t_0(0) =$ $- 10^{-1},$ $ -10^{-2},$
$-10^{-3},$ $ -10^{-4},$ and $ 0$. 
Before the formation of the naked singularity, the perturbations obey
the regularity conditions at the center. Each line in these figures 
displays this dependence if the radial coordinate is sufficiently small.
In this region, we can also see that all the variables grow according to 
power-laws on the time coordinate along the lines of $r=\mbox{const.}$
The asymptotic behavior of perturbations near the central naked
singularity is summarized as follows:
\begin{eqnarray}
  q \propto \Delta t ^{-2.1} r^4,~~ &k \propto \Delta t ^{-1.4} r^2, &~~k_{01}
    \propto \Delta t ^{-1.0} r^3, \nonumber \\ 
  \frac{\delta \rho}{\bar{\rho}} \propto \Delta t ^{-1.6} r^2, 
   ~~&V_1 \propto \Delta t ^{-0.4} r, &~~
   V_2 \propto \Delta t ^{-0.4} r^2,  \nonumber
\end{eqnarray}
where $\Delta t = t_0(0) - t$. 
On the time slice at $\Delta t =0$, perturbations behave as
\begin{eqnarray}
  q \propto r^{-0.09},~~ &k \propto r^{-0.74}, &~~k_{01} \propto
  r^{0.92}, \nonumber \\ 
  \frac{\delta \rho}{\bar{\rho}} \propto r^{-1.4}, 
   ~~&V_1 \propto r^{0.25}, &~~ V_2 \propto r^{1.3}.  \nonumber
\end{eqnarray}
On this slice $k$ and ${\delta \rho}/{\bar{\rho}}$ diverge and $q$
diverges weakly when they approach the central singularity. 
On the other hand, $k_{01}$ and $V_2$ go to zero and $V_1$ vanishes slowly.

\begin{figure}
  \begin{center}
    \leavevmode
    \epsfysize=250pt\epsfbox{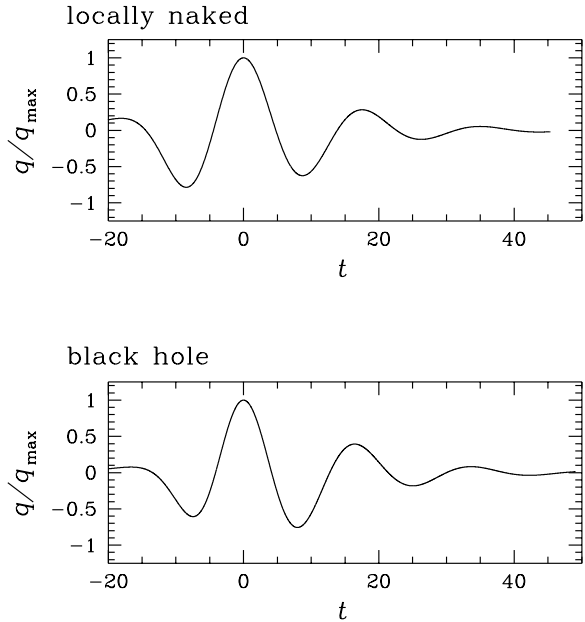}    
  \caption{Plots of perturbed variables  $q$  at constant
    circumferential radius $R=100$ in the locally naked and black hole cases.
    $q$ is normalized with respect to its maximum value, and the  
    origin of the time variable is adjusted to coincide with the time
    when $q$ is maximum.} 
 \label{fig:damp}
  \end{center}
 \end{figure}

In cases of a locally naked singularity and black hole formation, 
we expect to observe damped oscillation in the asymptotic region outside
the dust cloud, as in the odd parity case.  
The results are plotted in Fig.\ \ref{fig:damp}. These
figures show that damped oscillations are dominant. 
We read the frequencies and 
damping rates of these damped oscillations from Fig.\
\ref{fig:damp} and give them in terms of 
complex frequencies as $0.36+0.096i$ and $0.36+0.093i$ for locally naked
and black hole cases, respectively. These results agree well with the
fundamental quasi-normal frequency of the quadrupole mode
$(2M\omega = 0.74734 + 0.17792 i)$ \cite{Chandra}. \

\begin{figure}
  \begin{center}
    \leavevmode
    \epsfysize=250pt\epsfbox{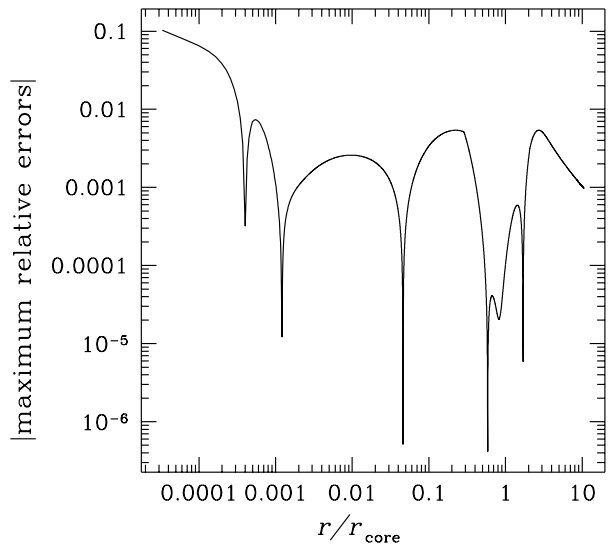}    
  \caption{Maximum relative errors on the last null slice.} 
 \label{fig:seido}
  \end{center}
 \end{figure}

The numerical accuracy of our calculations
was checked with the equations that were not used for
the derivation of Eqs.\ (\ref{evQ})--(\ref{evk01}), e.g., Eq.\
(\ref{02}). We define the maximum relative error $\cal{E}$ as
\begin{equation}
  \label{const}
  {\cal{E}} \equiv
  \frac{
    \begin{array}{lcr}
&&|-2\frac{R^3}{R'^8}X +2\frac{R^2}{R'^4}W
  +\frac{R^3}{R'^6}\left(6\frac{R''}{R'^2}-\frac{3}{R}
  -7\frac{\dot{R}}{R}+8\frac{\dot{R'}}{R'}\right)Z~~~~~~~~~~~\\
&&~~~~-\frac{R^3}{R'^7}\partial_{\tilde{r}}Z
+\frac{R^2}{R'^4}\left(4\frac{\dot{R}}{R}-6\frac{\dot{R'}}{R'}\right)\tilde{k}
  +16\pi\bar{\rho}V_2|
    \end{array}
  }{\Sigma |\mbox{each term of numerator}|}.
\end{equation}
We calculated this quantity on the last null surface
where the matter variable 
$V_2$ is obtained from the integration of Eq.\ (\ref{V2dot}) using a
method similar to that used for Eqs.\ (\ref{dotW})--(\ref{dotk}).
The results are displayed in Fig.\ \ref{fig:seido}.  Except the region of
small $r$ say, ($r < 3\times 10^{-4}$), this value is less than 0.01.
Both the numerator and denominator of Eq.\ (\ref{const}) vanish at the
center. Therefore it seems difficult to estimate the numerical errors from Eq.\
(\ref{const}) when $r$ is small. 

\section{Discussions}
\label{sec:discussion}
In this section we consider the physical interpretation of our
numerical results for even-parity perturbations. The divergence
behavior of the perturbations implies that 
the linear perturbation analysis near the Cauchy is invalid.
This fact shows that 
aspherical effects are important in the naked singularity formation. 

To consider where these effects are important and what would happen in
this region, we should discuss our results more carefully. 
The perturbations grow according to power-laws and 
diverge only at the Cauchy horizon. Therefore, except for the 
region  very near the Cauchy horizon, the perturbations are finite and
small when we 
choose sufficiently small initial values. 
This means that the central region can reach an extremely high density
before the breakdown of the linear perturbation analysis.
While in the region of spacetime
just before the Cauchy horizon, aspherical property becomes important for the
dynamics of the spacetime. Our results suggest that the Cauchy horizon
is unstable and that a singularity appears along it. 

The naked singularity of the LTB spacetime is considered as a massless
singularity. Gravitational waves, even if they have finite energy, would 
affect the naked singularity. To investigate this effect we should
consider back-reaction of the gravitational waves.

For the case of collapse that is not marginally bound, 
the condition of the appearance
of the central naked singularity is slightly different from that in the above
case \cite{Singh:1996tb,Jhingan:1996jb}, and hence there is a possibility 
that the behavior of perturbations in this case is different from that
in the 
marginally bound case. However, it is well known that 
the limiting behavior of the metric with
$t \rightarrow t_0(r)$ is common to all cases \cite{Landau}: 
\begin{equation}
  R \approx \left(\frac{9F}{4}\right)^{1/3}\left(t_0-t\right)^{2/3}
  ,~~~A \approx \left(\frac{2F}{3}\right)^{1/3} \frac{t_0'}{\sqrt{1+f}}
  \left(t_0-t\right)^{-1/3}.
\end{equation}
Thus we can conjecture that the results of the 
perturbation analysis for non-marginal collapse would be similar to 
the results for the marginally bound case.

\section{Summary}
\label{sec:summary}
We have studied the behavior of even-parity perturbations
in the LTB spacetime.
We have numerically solved the linearized Einstein equations for 
gauge-invariant variables  in the case of the quadrupole mode
and marginally bound background. 
We have constructed a numerical code which solves the perturbation
equations on an out-going single null coordinate. For the globally naked
case, the perturbed variables $q, k_{01}$ and the Weyl scalar $\Psi_4$ grow
as powers of $(u_0 - u)$ outside the dust cloud, 
where the power index is approximately $-0.88$. Then the Cauchy horizon
of this spacetime is unstable with respect to linear even-parity perturbations.
On the other hand, the
perturbed variable $k$ is finite just before the crossing of the Cauchy
horizon. The energy flux, which is proportional to the square of $k$, is
also finite. Therefore inhomogeneous aspherical dust collapse is
not expected to be a strong source of gravitational wave bursts.

We have investigated the asymptotic behavior of perturbations near a
central naked singularity. If the radial coordinate is sufficiently
small, the dependence on it is
determined by the regularity conditions at the center. Our numerical
results show this dependence. 
The time dependence is an inverse power-law in $\Delta t$. 
At the time of naked singularity formation, 
$q$, $k$ and ${\delta \rho}/{\bar{\rho}}$ diverge 
when they approach the central singularity, while 
$k_{01}$, $V_1$ and $V_2$ do not.

For the cases of locally naked and black hole formation, there appear
the damped oscillations outside the dust cloud. This is consistent with
the fundamental quasi-normal frequency of the quadrupole  mode
of a Schwarzschild black hole. 
 
\section*{Acknowledgements}
We would like to thank T. Nakamura for helpful and useful discussions.
We are also
grateful to H. Sato and colleagues in the theoretical astrophysics group
at Kyoto University for useful comments and encouragement.
This work was partly supported by Grants-in-Aid for Scientific
Research (No. 9204) and Creative Basic Research (No. 09NP0801)
from the Japanese Ministry of Education,
Science, Sport, and Culture.

\appendix
\section{Linearized Einstein Equations}
\label{sec:linear}
The linearized Einstein equations for the spherically symmetric
background presented by Gerlach and Sengupta are
\begin{eqnarray}
\label{le1}
2v^c\left(k_{ab|c}-k_{ca|b}-k_{cb|a}\right)
-\left[\frac{l(l+1)}{R^2}+G_c^{~c}+G_A^{~A}+2\cal{R}\right]k_{ab} &&\nonumber\\
-2g_{ab}v^c\left(k_{ed|c}-k_{ce|d}-k_{cd|e}\right)g^{ed}
+g_{ab}\left(2v^{c|d}+4v^cv^d-G^{cd}\right)k_{cd} && \nonumber\\
+g_{ab}\left[\frac{l(l+1)}{R^2}+\frac{1}{2}\left(G_c^{~c}+G_A^{~A}\right)
+\cal{R}\right]k_d^{~d}+2\left(v_ak_{,b}+v_bk_{,a}+k_{,a|b}\right)&&\nonumber\\
-g_{ab}\left[2{k_{,c}}^{|c}+6c^ck_{,c}-\frac{(l-1)(l+2)}{R^2}k\right]=
-16\pi T_{ab},&& 
\end{eqnarray}
\begin{eqnarray}
\label{le2}
    k_{,a}-{k_{ac}}^{|c}+{{k_c}^c}_{|a}-v_ak_c^{~c} &=& -16\pi T_a, 
\end{eqnarray}
\begin{eqnarray}
\label{le3}
  -\left({k_{,c}}^{|c}+2v^ck_{,c}+G_A^{~A}k\right)
+\left[{k_{cd}}^{|c|d}+2v^c{k_{cd}}^{|d}+2(v^{c|d}+v^cv^d)k_{cd}\right]
  && \nonumber\\
-g_{ab}\left[{{{k_c}^c}_{|d}}^{|d}+v^c{{k_d}^d}_{|c}
+{\cal{R}}k_c^{~c}-\frac{l(l+1)}{R^2}k\right] =-16\pi T^3, &&
\end{eqnarray}
\begin{eqnarray}
\label{le4}
  k_c^{~c} &=&-16\pi T^2,
\end{eqnarray}
where $\cal{R}$ is the Gaussian curvature of the 2-dimensional
submanifold $M^2$ spanned by $x^a$. Here
\begin{eqnarray}
  G_{ab} &\equiv& -2\left(v_{a|b}+v_av_b\right)+g_{ab}\left(2v_a^{~|a}
                +3v_av^a-\frac{1}{R^2}\right), \\
  G_A^{~A}  &\equiv& 2\left(v_a^{~|a}+v_av^a -{\cal{R}}\right).
\end{eqnarray}

For marginally bound LTB spacetime the linearized quadrupole Einstein
equations  are 
\begin{eqnarray}
\label{00}
    \frac{4}{R^2}q +\frac{1}{RR'}q' +\frac{\dot{R}}{R}\dot{q} &&   \nonumber\\
          -\frac{6}{R^2}{k} +\left(\frac{2}{RR'}-\frac{R''}{R'^3}\right)k' 
          -\left(2\frac{\dot{R}}{R}+\frac{\dot{R'}}{R'}\right)\dot{k}    
        +\frac{1}{R'^2}k'' &&   \nonumber\\ +2\left(\frac{\dot{R}}{R^2R'}
          -\frac{\dot{R}R''}{RR'^3} 
          +\frac{\dot{R'}}{RR'^2}\right){k_{01}}
          +2\frac{\dot{R}}{RR'^2}{k_{01}}' &=& -8 \pi \delta \rho, 
\end{eqnarray}
\begin{eqnarray}
  \label{01}
 -\frac{\dot{R}}{R}q' +\frac{R'}{R}\dot{q} 
   +\left(2\frac{\dot{R}}{R} -\frac{\dot{R'}}{R'}\right)k' +\dot{k'}  
   -\frac{3}{R^2}{k_{01}} &=& -8 \pi \bar{\rho} V_1,  
\end{eqnarray}
\begin{eqnarray}
  \label{11}
   2\frac{R'^2}{R^2}q  -\frac{R'}{R}q' 
     -\frac{R'^2\dot{R}}{R}\dot{q} +4\frac{R'^2\dot{R}}{R}\dot{k}  
                  +R'^2\ddot{k}     &&   \nonumber\\
    -2\frac{R'\dot{R}}{R^2}{k_{01}} -2\frac{R'}{R}\dot{k_{01}} 
     &=& 0,  
\end{eqnarray}
\begin{eqnarray}
  \label{02}
  -2\frac{\dot{R'}}{R'}q -\dot{q} +2\frac{\dot{R'}}{R'}{k} +2\dot{k} 
  +\frac{R''}{{R'}^3}{k_{01}}   -\frac{1}{R'^2}{k_{01}}'      
  &=& -16 \pi \bar{\rho} V_2,  
\end{eqnarray}
\begin{eqnarray}
  \label{12}
  q' +\dot{k_{01}} +\frac{\dot{R'}}{R'}{k_{01}} &=& 0,  
\end{eqnarray}
\begin{eqnarray}
  \label{22}
   \left(\frac{R^2R''}{R'^3} -2\frac{R}{R'}\right)q' 
                  -\left(2R\dot{R}+3\frac{R^2\dot{R'}}{R'}\right)\dot{q} 
                  -\frac{R^2}{R'^2}q'' -R^2\ddot{q}  & &  \nonumber\\
                    +4\left(R\dot{R}+\frac{R^2\dot{R'}}{R'}\right)\dot{k} 
                  +2R^2\ddot{k} 
  +2\left(\frac{RR''\dot{R}}{R'^3}-\frac{R\dot{R'}}{R'^2}\right)k_{01} & &  \nonumber\\
                  -2\frac{R\dot{R}}{R'^2}{k_{01}}'
                  +2\left(\frac{R^2R''}{R'^3}-\frac{R}{R'}\right)\dot{k_{01}}
                  -2\frac{R^2}{R'^2}\dot{k_{01}'} &=& 0,    
\end{eqnarray}
\begin{eqnarray}
  \label{z22}
  -k_{00} + \frac{1}{R'^2}k_{11} &=& 0. 
\end{eqnarray}
Here we have used Eq.\ (\ref{z22}) to eliminate $k_{11}$ in Eqs.\
(\ref{00})--(\ref{22}).

\section{Coefficients of Differential Equations}
\label{sec:evolve} 
The coefficients of Eqs.\ (\ref{dX})--(\ref{delq}) are
\begin{eqnarray}
  a_1 &=& \frac{1}{R}-\frac{\dot{R}}{R}-4\frac{R''}{R'^2}+\frac{5}{2}\frac{\dot{R}'}{R'},\\
  a_2 &=& 2\frac{R'^3}{R^2}\left(R\dot{R'}-R'\dot{R}\right),\\
  a_3 &=& 3\frac{R'^2\dot{R}^2}{R^2}-3\frac{R'\dot{R}\dot{R}'}{R},\\
  a_4 &=& -4\frac{R'^4\dot{R}^2}{R^3}+12\frac{R'^3\dot{R}\dot{R}'}{R^2}-8\frac{R'^2\dot{R}'^2}{R},\\
  a_5 &=& -3\frac{R'}{R}-\frac{1}{4}\frac{R'\dot{R}^2}{R}+\frac{1}{2}\dot{R}\dot{R}',
\\
  b_1 &=& -\frac{1}{R'^4},\\
  b_2 &=& -8\frac{\dot{R}}{R}+8\frac{\dot{R}'}{R'},\\
  b_3 &=& 0,\\
  b_4 &=& -7\frac{\dot{R}^2}{R^2}+28\frac{\dot{R}\dot{R}'}{RR'}-20\frac{\dot{R}'^2}{R'^2},\\
  b_5 &=& -\frac{3}{R'^3}-4\frac{\dot{R}}{R'^3}-\frac{9}{2}\frac{\dot{R}^2}{R'^3}+4\frac{R\dot{R}'}{R'^4}+4\frac{R\dot{R}\dot{R}'}{R'^4},
\\
  c_1 &=& -\frac{1}{R'^2},\\
  c_2 &=& 0,\\
  c_3 &=& -5\frac{\dot{R}}{R}+6\frac{\dot{R}'}{R'},\\
  c_4 &=& 0,\\
  c_5 &=& -\frac{1+\dot{R}}{R'},
\\  
  d_1 &=& 0,\\
  d_2 &=& 1,\\
  d_3 &=& 0,\\
  d_4 &=& 0,\\
  d_5 &=& \frac{R\left(1+\dot{R}\right)}{R'^3},
\\
  e_1 &=& \frac{1}{R},\\
  e_2 &=& 0,\\
  e_3 &=& 2\frac{R'}{R^2}\left(R'\dot{R}-R\dot{R}'\right),\\
  e_4 &=& 0,\\
  e_5 &=& 7\frac{R''}{R'}-3\frac{R'}{R}\left(1+\dot{R}\right)+5\dot{R}'.
\end{eqnarray}

\section{Power of Gravitational Radiation}
\label{sec:power}
In this appendix we calculate the radiated power of the gravitational
waves in an attempt to 
grasp the physical meaning of the gauge-invariant
quantities \cite{CPM}. \ To relate the perturbation of the metric to the 
radiated
gravitational power, it is useful to specialize to the radiation gauge,
in which the tetrad components $h_{(\theta)(\theta)}-h_{(\phi)(\phi)}$
and $h_{(\theta)(\phi)}$ fall off as $O(1/R)$, and all other tetrad
components fall off as $O(1/R^2)$ or faster. Note that in vacuum at
large distance, the spherically symmetric background metric is identical 
to the Schwarzschild solution, where hereafter we adopt the Schwarzschild
coordinates,  
\begin{equation}
  \label{Schmetric}
  ds^2 = -\left( 1-\frac{2M}{R}\right)d\tau ^2 +\left
  ( 1-\frac{2M}{R}\right)^{-1} dR^2 + R^2 \left( 
  d\theta^{2}+\sin^{2}\theta d\phi^{2}\right) . 
\end{equation}
The relation between the line elements Eq.\ (\ref{bgmetric}) and
Eq.\ (\ref{Schmetric}) is given by the transfer matrix 
\begin{eqnarray}
  d\tau &=& \frac{1}{1-(\partial_t R)_r^2} \{dt+(\partial_r R)_t
  (\partial_t R)_r
  dr\}, \\
  dR &=& (\partial_t R)_r dt + (\partial_r R)_t  dr.
\end{eqnarray}
In this gauge, the metric perturbations in Eq. (\ref{GS-pmetric}) behave
as 
\begin{eqnarray}
  h_{ab} &=& O\left(\frac{1}{R^2}\right), \\
  h_a &=& O\left(\frac{1}{R}\right), \\
  K &=& O\left(\frac{1}{R^2}\right), \\
  G &=& \frac{g(\tau-R_*)}{R} + O\left(\frac{1}{R^2}\right), 
\end{eqnarray}
where 
\begin{equation}
  R_* = R + 2M \ln \left( \frac{R}{2M} - 1\right) + \mbox{const},
\end{equation}
and the out-going wave condition is respected.
Then, the gauge-invariant metric perturbations (\ref{gi-kab}) and
(\ref{gi-k}) are calculated as 
\begin{eqnarray}
  k_{\tau \tau} &=& g^{(2)}R+O(1), \\
  k_{\tau R} &=& -g^{(2)}R+O(1), \\
  k_{RR} &=& g^{(2)}R+O(1), \\
  k &=& -g^{(1)}+O\left(\frac{1}{R}\right),
\end{eqnarray}
where $g^{(1)}$ denotes the first derivative of $g$ 
with respect to its argument.

In this radiation gauge, the radiated power $P$ per unit solid angle is
given by the formula derived by Landau and Lifshitz
\cite{Landau} from their stress-energy pseudo-tensor,
\begin{equation}
  \frac{dP}{d\Omega}=\frac{R^2}{16\pi}\left[\left(\frac{\partial
  h_{(\theta)(\phi)}}{\partial \tau}\right)^2
  +\frac{1}{4}\left(\frac{\partial h_{(\theta)(\theta)}}{\partial
  \tau}-\frac{\partial h_{(\phi)(\phi)}}{\partial \tau}\right)^2\right].
\end{equation}
For the axisymmetric mode, i.e. $m=0$, the above formula is reduced to 
\begin{equation}
  \frac{dP}{d\Omega}=\frac{1}{64\pi}(g^{(1)})^2A_l (\theta),  
\end{equation}
where
\begin{equation}
  A_l (\theta) \equiv \frac{2l+1}{4\pi}\sin^4\theta\left(\frac{d^2P_l(\cos\theta)}{(d\cos\theta)^2}\right)^2.
\end{equation}
By using the gauge-invariant quantities and integrating over 
all solid angles, the formula for the power of the gravitational radiation
is obtained in the following form:
\begin{eqnarray}
  \frac{dP}{d\Omega}&=&\frac{1}{64\pi}k^2 A_l (\theta),\\
  P &=& \frac{1}{64\pi} B_l k^2 ,  
\end{eqnarray}
where
\begin{equation}
  B_l \equiv \frac{(l+2)!}{(l-2)!}.
\end{equation}

\end{document}